\documentclass[preprint]{aastex}
\usepackage{epsfig}
\def\comma{,~}
\begin{document}

\title{Evidence for TeV Emission from GRB 970417a}

\author{
R.~Atkins,\altaffilmark{1}
W.~Benbow,\altaffilmark{2}
D.~Berley,\altaffilmark{3,10}
M.L.~Chen,\altaffilmark{3,11}
D.G.~Coyne,\altaffilmark{2}
B.L.~Dingus,\altaffilmark{1}
D.E.~Dorfan,\altaffilmark{2}
R.W.~Ellsworth,\altaffilmark{5}
D.~Evans,\altaffilmark{4}
A.~Falcone,\altaffilmark{6}
L.~Fleysher,\altaffilmark{7}
R.~Fleysher,\altaffilmark{7}
G.~Gisler,\altaffilmark{8}
J.A.~Goodman,\altaffilmark{3}
T.J.~Haines,\altaffilmark{8}
C.M.~Hoffman,\altaffilmark{8}
S.~Hugenberger,\altaffilmark{4}
L.A.~Kelley,\altaffilmark{2}
I.~Leonor,\altaffilmark{4}
M.~McConnell,\altaffilmark{6}
J.F.~McCullough,\altaffilmark{2}
J.E.~McEnery,\altaffilmark{1}
R.S.~Miller,\altaffilmark{8,6}
A.I.~Mincer,\altaffilmark{7}
M.F.~Morales,\altaffilmark{2}
P.~Nemethy,\altaffilmark{7}
J.M.~Ryan,\altaffilmark{6}
B.~Shen,\altaffilmark{9}
A.~Shoup,\altaffilmark{4}
C.~Sinnis,\altaffilmark{8}
A.J.~Smith,\altaffilmark{9,3}
G.W.~Sullivan,\altaffilmark{3}
T.~Tumer,\altaffilmark{9}
K.~Wang,\altaffilmark{9}
M.O.~Wascko,\altaffilmark{9}
S.~Westerhoff,\altaffilmark{2}
D.A.~Williams,\altaffilmark{2}
T.~Yang,\altaffilmark{2}
G.B.~Yodh\altaffilmark{4} \\
(The Milagro Collaboration)}

\altaffiltext{1}{University of Utah, Salt Lake City, UT\,84112, USA}
\altaffiltext{2}{University of California, Santa Cruz, CA\,95064, USA}
\altaffiltext{3}{University of Maryland, College Park, MD\,20742, USA}
\altaffiltext{4}{University of California, Irvine, CA\,92697, USA}
\altaffiltext{5}{George Mason University, Fairfax, VA\,22030, USA}
\altaffiltext{6}{University of New Hampshire, Durham, NH\,03824, USA}
\altaffiltext{7}{New York University, New York, NY\,10003, USA}
\altaffiltext{8}{Los Alamos National Laboratory, 
                 Los Alamos, NM\,87545, USA}
\altaffiltext{9}{University of California, Riverside, CA\,92521, USA}
\altaffiltext{10}{Permanent Address: National Science Foundation, Arlington, VA\
,22230, USA}
\altaffiltext{11}{Now at Brookhaven National Laboratory, Upton, NY\,11973, USA}

\begin{abstract}
Milagrito, a detector sensitive to very high energy gamma rays,
monitored the northern sky from February 1997 through May 1998. With a
large field of view and a high duty cycle, this instrument was well
suited to perform a search for TeV gamma-ray bursts (GRBs).  We
report on a search made for TeV counterparts to GRBs observed by
BATSE. BATSE detected 54 GRBs within the field of view of Milagrito
during this period.  An excess of events coincident in time and space
with one of these bursts, GRB 970417a, was observed by Milagrito. The
excess has a chance probability of $2.8 \times 10^{-5}$ of being a
fluctuation of the background. The probability for observing an excess
at least this large from any of the 54 bursts is $1.5 \times 10^{-3}$.
No significant correlations were detected from the other bursts.
\end{abstract}

\section{Introduction}

Gamma-ray bursts were discovered over 30 years ago~\cite{klebsdell73}.
Although thousands of GRBs have been observed the physical processes
responsible for them are still unknown.  The understanding of these
objects was greatly enhanced by results from the Compton Gamma Ray
Observatory, which contained experiments sensitive to photons from 50
keV to 30 GeV. One of these experiments, BATSE, a wide field
instrument sensitive to gamma rays from 50 keV to above 300
keV~\cite{batse}, has detected several thousand GRBs.
EGRET~\cite{egret} detected 7 GRBs with photon energies
ranging from 100 MeV to 18 GeV~\cite{egretgrb}.  No high energy cutoff
above a few MeV has been observed in any GRB spectrum, and emission up
to TeV energies is predicted in several
models~\cite{dermer99,pilla98,totani98a,meszaros94}.

Very high energy gamma-ray emission may not be observable for sources at
redshifts much greater than 0.5 because of pair production with
infrared extragalactic background photons~\cite{jelley66,gould66}.
Recent observations of lower-energy afterglows associated with several
GRBs have allowed the measurement of 9 redshifts,
either by measuring the spectrum of the optical afterglow, or by
measuring the spectrum of the putative host galaxy. These redshifts cover 
a range between 0.4 and 3.4, and imply that the distribution of intrinsic
luminosities is broad~\cite{afterglows}. This suggests that the
intensity of TeV gamma-ray emission from a GRB (which
requires a relatively nearby source) may not be well correlated with
the intensity of the sub-MeV emission detected by BATSE.

At energies greater than 30 GeV, gamma-ray fluxes from most
astrophysical sources become too small for current satellite-based
experiments to detect because of their small sensitive areas.  Only
ground-based experiments \cite{hoffman99, ong98, weekes99} have areas
large enough to detect these sources. These instruments detect the
extensive air showers produced by the high energy photons in the
atmosphere, thus giving them a much larger effective area at high
energies.  These showers can be observed by detecting the Cherenkov
light emitted by the cascading relativistic particles as they traverse
the atmosphere, or by detecting the particles which reach ground
level.

TeV gamma-ray emission from several astrophysical sources has been
detected using atmospheric Cherenkov telescopes.  These instruments
have extremely large collection areas ($\sim 10^{5}$ m$^{2}$) and good
hadronic rejection. Unfortunately, they have relatively narrow fields of
view (a few degrees) and can operate only on dark clear nights,
resulting in a low duty cycle. They are therefore ill suited to
search for transient sources such as GRBs.  Searches for GRBs at
energies above 300 GeV have been made by slewing these telescopes
within a few minutes of the notification of the GRB
location~\cite{connaughton97}. No detections have been reported.
However, because of the narrow field of view, coupled with the delay in
slewing to the correct position, there have not been any prompt TeV
gamma-ray observations at the GRB location.

At energies greater than 10 TeV, the Tibet collaboration reported a
possibly significant deviation of the probability distribution from
background, for the superposition of all the bursts within their field
of view.  However, no single burst showed a convincing
signal~\cite{amenomori96}.  Two GRBs occurred within the field of view
of the HEGRA AIROBICC Cherenkov array. One very long duration burst
showed an excess over background from a direction not entirely
consistent with the sub-MeV emission, so this was not claimed as a
firm detection~\cite{padilla98}.

Milagro, a new type of TeV gamma-ray observatory with a field of
view greater than one steradian and a high duty cycle, began operation
in December 1999 near Los Alamos, New Mexico.  A prototype detector,
Milagrito~\cite{atkins99b}, operated from February 1997 to May 1998,
during which 54 GRBs detected by BATSE were within 45$^{\circ}$ of
zenith of Milagrito. This paper reports on the search for TeV
gamma-ray emission from these 54 gamma-ray bursts, but
concentrates more specifically on GRB 970417a.

\section{The Milagrito Detector}

Milagrito consisted of a planar array of 228 8-inch photomultiplier
tubes (PMTs) submerged in a light-tight water
reservoir~\cite{atkins99b}.  The PMTs were located on a square grid
with 2.8 m spacing, covering a total area of 1800 m$^2$.  Data were
collected at water depths of 0.9, 1.5 and 2.0 m above the PMTs. The
PMTs detected the Cherenkov light produced as charged shower particles
traversed the water.  The abundant gamma rays in the air shower
interact with the water via pair production and Compton scattering to
produce additional relativistic charged particles, increasing the
Cherenkov light yield.  The continuous medium and large Cherenkov
angle (41$^\circ$ ) result in the efficient detection of shower
particles incident on the reservoir with the array of PMTs.
Simulations show that Milagrito was sensitive to showers produced by
primary gamma rays with energies as low as $\sim$100 GeV. The
relative arrival times of the shower front at the PMTs were used to
reconstruct the direction of the incoming air shower.  The trigger
required $>$100 PMTs to register at least one photoelectron within a
300 ns time window.  Events were collected at a rate of
$\sim$300~s$^{-1}$; almost all of these triggers were caused by the
hadronic cosmic-ray background. The capability of Milagrito to detect
TeV gamma rays was demonstrated by the observation of the active
galaxy Markarian 501 during its 1997 flare~\cite{atkins99a}. The
instrument had an angular resolution of about $1^{\circ}$.

\section{Observations and Results}

A search was conducted in the Milagrito data for an excess of events, 
above those due to the background of cosmic rays, coincident with BATSE
GRBs. Only bursts within 45$^{\circ}$ of zenith of Milagrito were
considered because the sensitivity of Milagrito fell rapidly with
increasing zenith angle.  For each burst, a circular search region on
the sky was defined by the BATSE 90\% confidence interval, which
incorporates both the statistical and systematic position
errors~\cite{briggs99}.  The search region was tiled with an array of
overlapping $1.6^{\circ}$ radius bins spaced $0.2^{\circ}$ apart in RA
and DEC. This radius was appropriate for the measured angular
resolution of Milagrito~\cite{atkins99a,atkins99b}.  The number of
events falling within each of the $1.6^{\circ}$ bins was summed for
the duration of the burst defined by the T90 interval reported by
BATSE.  This time period is that in which the BATSE fluence rose from
5\% to 95\% of its total. T90 was chosen, {\it a priori}, because the
EGRET detections were much more significant during T90 than during
longer time intervals~\cite{hurley94}.

For each GRB, the angular distribution of background events on the sky
was characterized using two hours of data surrounding each burst. This
distribution was then normalized to the number of events ($N_{T90}$)
detected by Milagrito over the entire sky during T90. The
resulting background data were also binned in the same $1.6^{\circ}$
overlapping bins as the initial data.  Each bin in the actual data was
compared to the corresponding bin in the background map. The Poisson
probability of a background fluctuation giving rise to an excess at
least as large as that observed was calculated.  The bin with the
lowest such probability was then taken as the most likely position of
a very high energy gamma-ray counterpart to that particular BATSE
burst.

\begin{figure}[tbh!]
\centerline{\epsfig{figure=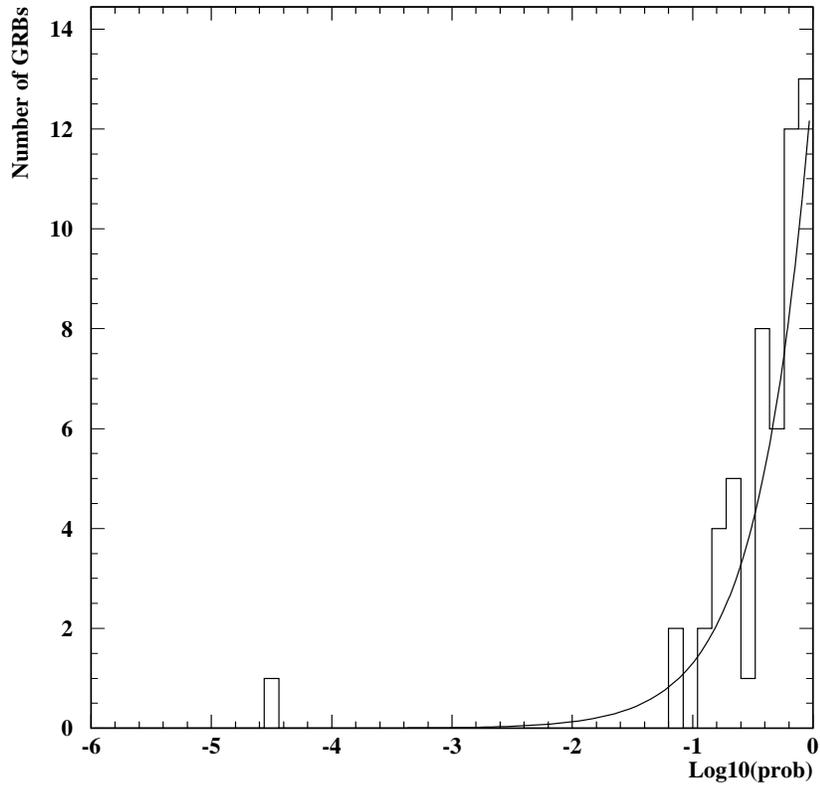,width=12cm}}
\caption{The distribution of probabilities, that the observed excess
number of events at the candidate TeV position was a background
fluctuation, for each of the 54 bursts. The curve indicates the 
expected distribution of probabilities for a sample drawn from background.
The entry at -4.5 corresponds to GRB 970417a.}
\label{fig:prob_mil}
\end{figure}

The chance probability of obtaining at least the observed significance
anywhere within the entire search region was determined by Monte Carlo
simulations using the following procedure. For each burst a set of
simulated signal maps was obtained by randomly drawing $N_{T90}$
events from the background distribution.  These maps were searched, as
before, for the most significant excess within the search region
defined by the BATSE 90\% confidence interval. The probability after
accounting for the size of the search region is given by the ratio of
the number of simulated data sets with probability less than that
observed in the actual data to the total number of simulated data
sets. The distribution of the chance probabilities obtained by this
method for the 54 GRBs is given in Figure~\ref{fig:prob_mil}. Details
of a somewhat different analysis, which yields consistent results with
those reported here, as well as more detailed results from the other
53 bursts, will be described elsewhere~\cite{isabelthesis}.

\begin{figure}[tbh!]
\centerline{\epsfig{figure=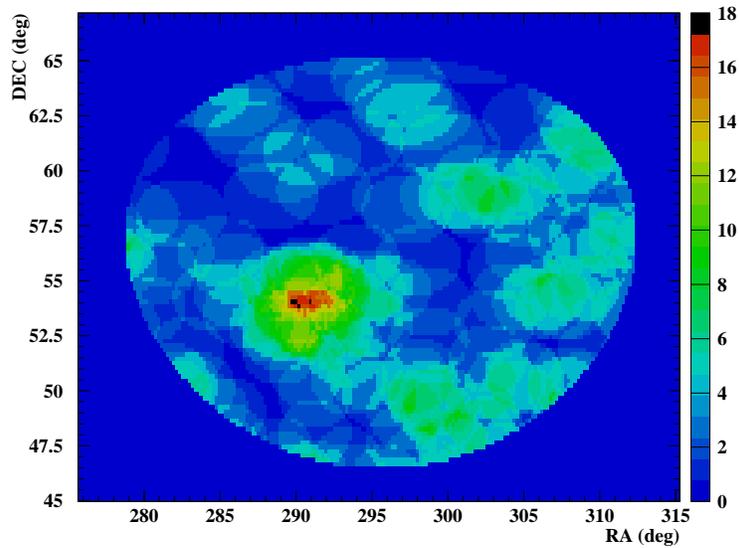,width=10cm}}
\caption{Number of events recorded by Milagrito during T90 in 
overlapping 1.6$^\circ$ radius bins in the vicinity of GRB 970417a. }
\label{fig:t90sky}
\end{figure}

One of these bursts, GRB 970417a, shows a large excess above
background in the Milagrito data.  The BATSE detection of this burst
shows it to be a relatively weak burst with a fluence in the 50--300
keV energy range of $1.5 \times 10^{-7}$ ergs/cm$^2$ and T90 of 7.9
seconds.  BATSE determined the burst position to be
RA~$=295.7^{\circ}$, DEC~$=55.8^{\circ}$.  The low BATSE fluence
results in a large positional uncertainty of $6.2^{\circ}$
(1$\sigma$).  The resulting search region for TeV emission has a
radius of $9.4^{\circ}$.  The $1.6^{\circ}$ radius bin with the
largest excess in the Milagrito data is centered at RA~$=
289.9^{\circ}$ and DEC~$= 54.0^{\circ}$, corresponding to a Milagrito
zenith angle of $21^{\circ}$. This location is consistent with the
position determined by BATSE.  The uncertainty in the candidate
location is approximately $0.5^{\circ}$ (1$\sigma$), much better than
the BATSE uncertainty. Figure~\ref{fig:t90sky} shows the number of
counts in this search region for the array of $1.6^{\circ}$ bins. The
bin with the largest excess has 18 events with an expected background
of $3.46 \pm 0.11$ (statistical error based on the background
calculation method used). The Poisson probability for observing an
excess at least this large due to a background fluctuation is $2.9
\times 10^{-8}$. The probability of such an excess or greater anywhere
within the search region for this burst was found by the Monte Carlo
simulation described above to be $2.8 \times 10^{-5}$ (see
Figure~\ref{fig:prob_mil}). For 54 bursts, the chance probability of
background fluctuating to at least the level observed for GRB 970417a
for at least one of these bursts is $1.5\times 10^{-3}$. The
individual events contributing to this excess were examined. The
distributions of the number of tubes hit per event and the shower
front reconstructions were consistent with those from other
shower events.  There is no evidence that the detector was malfunctioning
during the burst analysis time period.

\begin{figure}[tbh!]
\centerline{\epsfig{figure=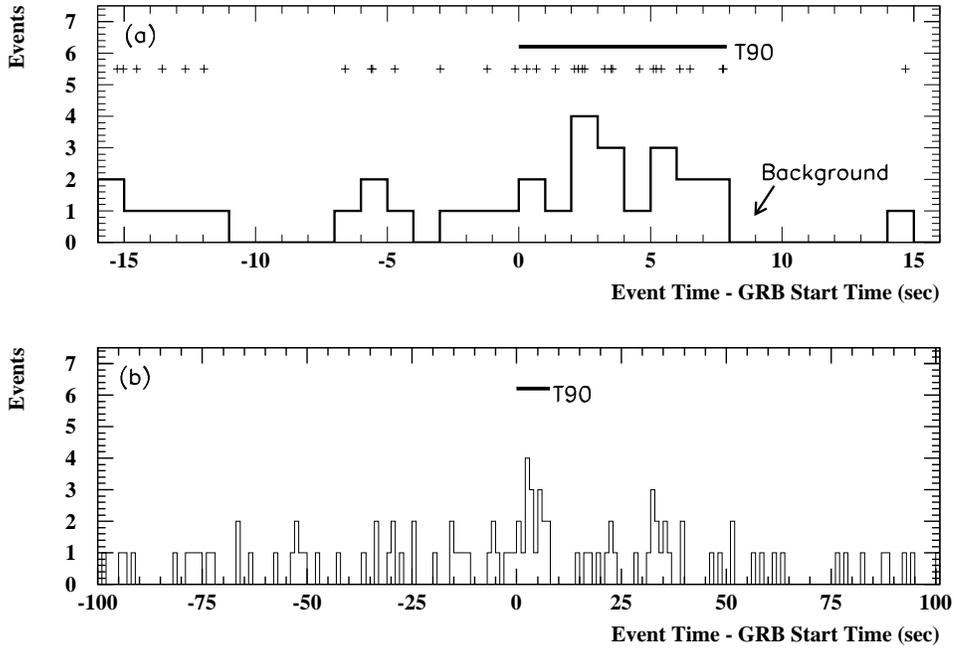,width=14cm}}
\caption{
GRB 970417a: (a) The crosses indicate the arrival time of events from
within a 1.6$^\circ$ radius of the candidate TeV counterpart for $\pm$15 s
around the start of T90. The histogram shows the same data binned in 1
second intervals. (b) The Milagrito data integrated in 1 sec intervals
for $\pm$100 s around the start of T90 (13:53:35.689 UT).}
\label{fig:lc}
\end{figure}

Although the initial search was limited to T90, upon identifying GRB
970417a as a candidate, longer time intervals were also
examined. EGRET observed longer duration GeV emission~\cite{hurley94},
and TeV afterglows are predicted by several
models~\cite{meszaros94,totani98b}. A search for TeV gamma rays
integrated over time intervals of one hour, two hours and a day after
the GRB start time did not show any significant excesses.  Histograms
of shorter time intervals, where the data are binned in intervals of
one second, are shown in Figure~\ref{fig:lc}.  An analysis of the data
also revealed no statistically significant evidence for TeV
after-flares.

\section{Discussion}

\begin{figure}[tbh!]
\centerline{\epsfig{figure=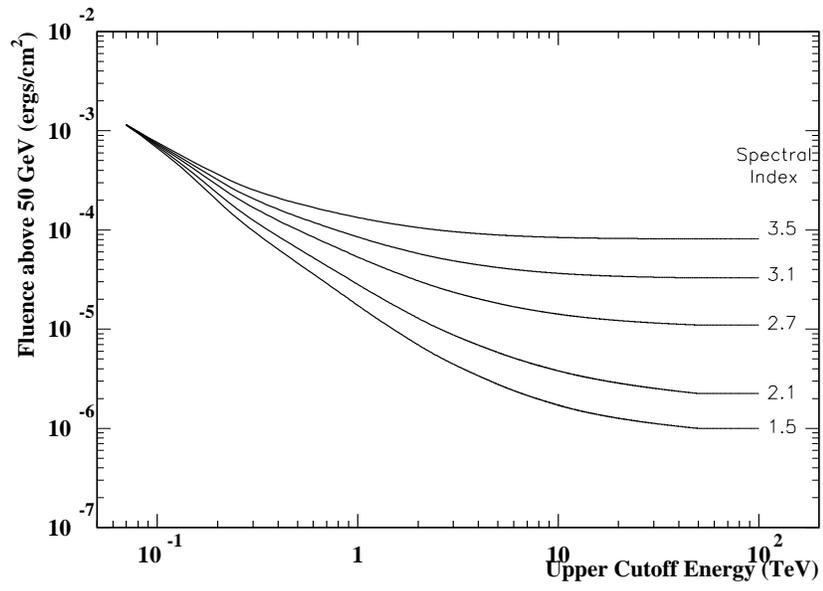,width=12cm}}
\caption{The implied fluence ($>$ 50 GeV) of very high energy  emission 
from GRB 970417a 
as a function of high-energy cutoff for five assumed differential spectral 
indices. }
\label{fig:ergs}
\end{figure}

If the observed excess of events in Milagrito is indeed associated
with GRB 970417a, then it represents the highest energy photons yet
detected from a GRB.  The energy spectrum and maximum energy
of emission are difficult to determine from Milagrito data.  The small
size of the pond compared to the lateral extent of typical air showers,
along with the poor ability of this instrument to measure the amount
of energy deposited in the pond, make the estimation of shower energy
on an event by event basis nearly impossible.  The very high energy
fluence implied by this observation depends on the spectrum and upper
energy cutoff of the emission, which Milagrito is unable to determine.
Monte Carlo simulations of gamma-ray-initiated air showers show
that the effective area of Milagrito increases slowly with energy, so
that the energy threshold is undefined~\cite{atkins99b}.  However,
Milagrito had very little sensitivity below 100 GeV, so this
observation indicates the emission of photons with energies
greater than a few hundred GeV from GRB 970417a.
Figure~\ref{fig:ergs} shows the implied fluence of this observation
above 50 GeV as a function of upper cutoff energy for several assumed
differential power-law spectra.  The observed cosmic-ray event rate
agrees well with the rate predicted by simulations~\cite{atkins99a}
implying that the systematic error on the energy scale for Milagrito
is $<$30\%.

Several studies \cite{salamon98,primack99} find that the opacity due
to pair production for $>$200 GeV gamma rays exceeds one for
redshifts larger than $\sim$0.3.  Thus, if Milagrito has detected high
energy photons from GRB 970417a, it must be a relatively nearby
object.  The observed excess implies a fluence above 50 GeV between
$10^{-3}$ and $10^{-6}$ ergs/cm$^{2}$ and the spectrum must extend to
at least a few hundred GeV.  The very high energy gamma-ray fluence ($>50$ GeV)
inferred from this result is at least an order of magnitude greater
than the sub-MeV fluence.

To summarize, an excess of events with chance probability $2.8 \times
10^{-5}$ coincident both spatially and temporally with the BATSE
observation for GRB 970417a was observed using Milagrito. The chance
probability that an excess of at least this significance would be
observed from the entire sample of 54 bursts is $1.5 \times 10^{-3}$.
If the observed excess coincident with GRB 970417a is not an unlikely
fluctuation of the background, then a GRB
bright at TeV energies has been identified.  A search for other
coincidences with BATSE will be continued with
the current instrument, Milagro, which has significantly increased
sensitivity to GRBs between 0.1 and 10 TeV.

\acknowledgments

Many people helped bring Milagrito to fruition.  In particular, we
acknowledge the efforts of Scott DeLay, Neil Thompson and Michael Schneider. 
This work was supported in part by the National Science Foundation, the U. S.
Department of Energy (Office of High Energy Physics and Office of Nuclear
Physics), Los Alamos National Laboratory, the University of California, and
the Institute of Geophysics and Planetary Physics.                          

\end{document}